\documentclass{article}

\usepackage[preprint]{nips_2018}

\usepackage[utf8]{inputenc} 
\usepackage[T1]{fontenc}    
\usepackage{hyperref}       
\usepackage{url}            
\usepackage{booktabs}       
\usepackage{amsfonts}       
\usepackage{nicefrac}       
\usepackage{microtype}      
\usepackage{graphicx}

\title{Constructing an olfactory perceptual space and predicting percepts from molecular structure}

\author{
  Daniel Kepple and Alexei Koulakov\\
 Cold Spring Harbor Laboratory\\
  Cold Spring Harbor, NY 11724 \\
  \texttt{dkepple@cshl.edu} \\
}

\begin{document}

\maketitle

\begin{abstract}
 Given the structure of a novel molecule, there is still no one who can reliably predict what odor percept that molecule will evoke. The challenge comes both from the difficulty in quantitatively characterizing molecular structure, and the inadequacy of language to fully characterize olfactory perception. Here, we present a novel approach to both problems. First, we avoid explicit characterization of molecular structure by using a similarity score for each molecular pair, derived from comparing the molecular structures directly. We show that this method is advantageous to conventional methods, improving predictions without relying on preexisting knowledge of chemical descriptors. Second, we generate a perceptual space, in which a molecule’s location defines its percept. We show that from a molecule’s neighbors in this space alone, we are able to reproduce all perceptual descriptors of that molecule. We propose that predicting olfactory percept from structure can be rethought of as predicting a molecule’s location in this perceptual space. This suggestion provides a framework for understanding and predicting human smell percepts.
\end{abstract}

\section{Introduction}

The relationship between a molecule’s chemical structure and the human olfactory percept that molecule will evoke has long remained mysterious. This is despite many years of experiment and analysis \citep{sell}. A fundamental challenge in this endeavor is to understand variation in molecular structure \citep{alex}. Using intuition or knowledge from chemistry to categorize and relate molecules, such as with functional group counts, has proved insufficient to reproduce percepts (\citep{laska}\citep{boes}). 

From recent research to visualize olfactory receptor responses to an ensemble of molecules \citep{joel}, we now appreciate the promiscuity and complexity of the olfactory receptor. Molecules that look similar to a researcher may elicit very different responses in a receptor, and vice versa. For this reason, modern attempts to characterize molecular structure have focused on machine learning approaches, hoping that a systematic analysis will capture what our intuition misses \citep{sobel}\citep{kerm}\citep{zarz}\citep{snitz}\citep{voss}. 

Any algorithm’s success, however, still hinges upon what information it is provided. Therefore, researchers have used every possible physicochemical measure imagined by chemists, collected into a software called Dragon 6 \citep{dragon}. These quantitative descriptors vary from simple functional group counts to normalized eigenvalue sums of the powers of connectivity matrices. It is often unclear how a particular property might be recognized by a receptor, or if the full structure of a molecule is truly encapsulated in these properties. What is clear, however, is that predictions made using these properties leave room for improvement \citep{sobel}\citep{snitz}\citep{voss}. 

Similarly, machine learning methods are limited by the ability of perceptual descriptors to characterize human olfactory perception. The perception of a smell is usually presented as a group of words such as “flowery,” “sweet,” or “rotten” \citep{acree}\citep{good}\citep{arctander}. The relationship or higher order organization between these coarse descriptors is unknown. Then, to make any percept prediction, a separate predictor must be built for each percept. These individual percept predictors necessarily use less data and ignore information about other percepts that could be critical in determining which molecular features are selected by olfaction. 

We present a novel approach for characterizing the olfactory molecular input space, as well as a new framework for understanding the space of human olfactory perception. We characterize molecules by direct comparison of their structure, requiring no prior knowledge of physicochemical properties. We stich together a perceptual space from local relationships between percepts, finding a low dimensional perceptual space while still retaining resolution of the full single-percept set. We show that our alignment-based method is advantageous to Dragon-based methods both in its parsimonious representation of molecular space and its ability to predict dimensions of perceptual space and individual percepts.  

\section{Results}
\label{gen_inst}

\subsection{Olfactory Perceptual Space}

\subsubsection{Databases}

We used five independent odorant datasets in our analysis: Flavornet \citep{acree}, GoodScents \citep{good}, Arctander\citep{arctander}, Vosshall\citep{voss}, and Dravnieks \citep{dravnieks}. Each dataset has a different number of molecules and descriptor system. Table 1 describes basic statistics of each dataset.

\begin{table}
  \caption{Human Olfactory Perceptual Datasets}
  \label{sample-table}
  \centering
  \begin{tabular}{llll}
    \toprule
    Name     & Number of Molecules     & Descriptors & Rating System ($\mu$m) \\
    \midrule
    Flavornet & 750 & 197 words  & Binary     \\
    GoodScents     & 3554 & 580 words   & Binary        \\
    Arctander     & 3102   & Prose    & N/A   \\
    Vosshall & 472l  & 21 words        & 1-100 \\
    Dravnieks     & 146 & 144  words       & Percentage of participants in agreement    \\
    \bottomrule 
  \end{tabular}
\end{table}

\subsubsection{Constructing a perceptual space}

Three of these databases use qualitative descriptors to label molecules: Flavornet, Arctander, and GoodScents.  These descriptors are highly sparse, making predictions from a high dimensional molecular space difficult. The other two datasets, Dravnieks and Vosshall, avoid this issue by averaging the responses of groups of participants to arrive at a quantitative description. This quantitative description forms the basis of a perceptual space, in which distance between molecules is the distance between their rating scores. 

For the three qualitative datasets, we developed a novel method for constructing a perceptual space. We connect all molecules of a database into a graph. In this graph, each node is a molecule and each weighted edge between two molecules is calculated based on the overlap between those two molecules’ percepts. In particular, we calculate each edge using the equation \begin{math}
E_{ij}=\left | \frac{O_{ij}}{\sqrt{O_{ii}O_{jj}}}-1 \right |
\end{math} where \begin{math} O_{ij} \end{math}  is the number of overlapping percepts between molecule \begin{math} i \end{math} and \begin{math} j \end{math}. 

Using this weighted overlap graph, we compute geodesic distance between all molecule pairs and embed with ISOMAP to  approximate a human olfactory perceptual space [Figure 1] \citep{iso}. For the Arctander dataset, which describes molecules in prose, we searched each molecule’s description for the 580 percepts of GoodScents which were then used to construct the overlap matrix. 

\begin{figure}
  \centering
\includegraphics[scale=.18]{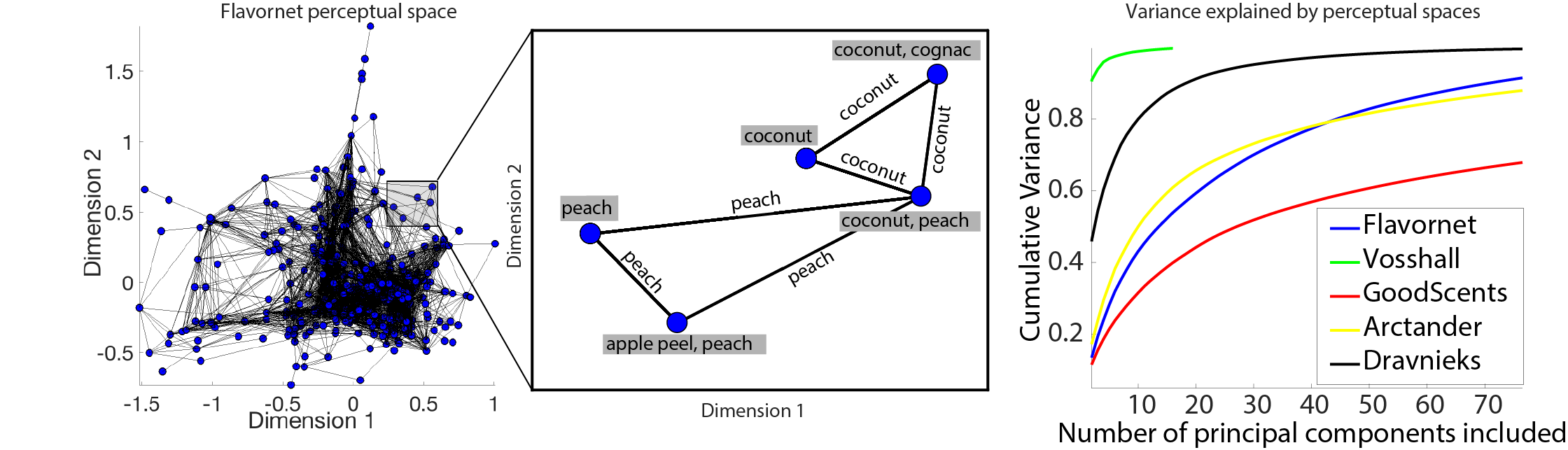}
  \caption{Perceptual spaces. (Left) Flavornet embedded using ISOMAP, with zoom-in to show graph scheme. (Right) Variance explained by dimensions of perceptual space for all five datasets}
\end{figure}

\subsubsection{Dimensions of perceptual space}

Given that each of the five datasets contain different molecules, descriptors, and acquisition techniques, we investigated whether or not the datasets were consistent.  

First, we compared the dimensionality of each dataset’s percept space from their principal component cumulative variance curves [Figure 1]. The general trend is that dimensionality increases with number of percepts. Notably, however, the Vosshall dataset, recently introduced for the DREAM olfaction challenge, is essentially one dimensional. This is mostly a result of the high variance of the “pleasantness” percept relative to other percepts. It is also possible that the untrained participants used in this study were not confident with the other descriptors. This is supported by the comparison of shared molecules using shared percepts across all datasets. In particular, comparing “floral” and “fruity” using only shared molecules yields high correlation between all dataset pairs except those including the Vosshall dataset.

We also note that the dimensionality of our graph perceptual spaces seems to be inflated compared to the Dravniek’s dataset, even after accounting for differences in perceptual descriptors. As our procedure may add high dimensional noise, we looked for which dimensions are most salient for recovering the original qualitative descriptors. 
To recover a molecule’s perceptual descriptors, we devised a simple score for each descriptor based on the percepts of neighboring molecules in perceptual space. Specifically, for each percept, we sum the distance of all molecules containing that percept from the test molecule with exponential decay controlled by a parameter lambda as is shown in the equation: \begin{math}S_{mp}=\sum_{n\neq m}P_{pn}e^{-\lambda D_{mn}}\end{math}. Here  \begin{math} S_{mp} \end{math} is the score given to percept  \begin{math} p \end{math} for molecule  \begin{math} m \end{math},  \begin{math} P_{pn} \end{math} is the percept matrix which is 1 if molecule  \begin{math} n \end{math} has percept  \begin{math} p \end{math}, and  \begin{math} D_{mn} \end{math} is the distance between the two molecules. An example of a sorted percept score for a test molecule with a tuned lambda is shown in [Figure 2]. As is intuitive from the example score, we predict both the number of percepts describing a molecule and the identity of those percepts by sorting a molecule’s percept scores and selecting the top  \begin{math} k \end{math} percepts, where  \begin{math} k_{m}=\max_{p}(\left | S_{mp}-S_{m(p+1)} \right |) \end{math}  . 

In Figure 2, we show the results of this recovery while iteratively increasing the number of perceptual dimensions used calculating the distance matrix \begin{math} D \end{math}  for Flavornet. Remarkably, all percepts can be recovered without error for the majority of molecules using only six dimensions. In the right panel of Figure 2, we calculate the F measure, or the harmonic mean of precision and recall, for the first 10 dimensions of all three graph perceptual spaces. Percept recovery performance in all datasets begin to plateau around 6D, with 91\% of molecules in Flavornet recovering their percepts with a maximum of one error; 86\% for Arctander; and 84\% of Goodscents. 

\begin{figure}
  \centering
\includegraphics[scale=.22]{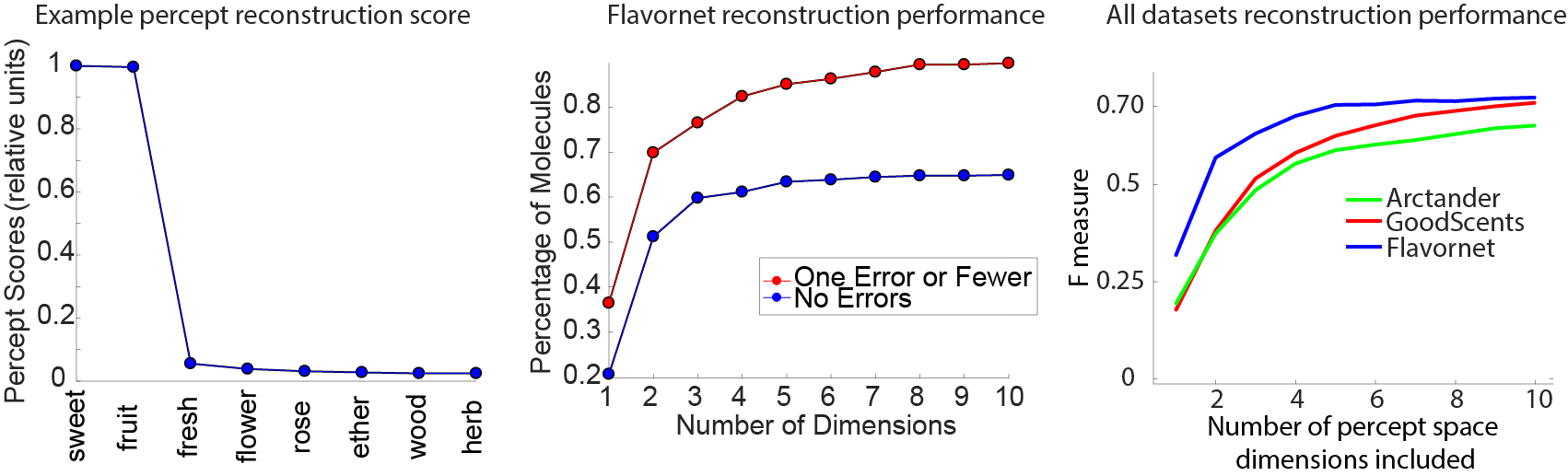}
  \caption{Perceptual descriptor recovery from perceptual space. (Left) Example percept recovery score for a molecule in Flavornet. (Center) Flavornet reconstruction performance. (Right) Average F measure between recovered and true of percepts across all molecules in all three graph perceptual spaces. }
\end{figure}

To understand what these six perceptual dimensions correspond to, we scored each percept for each dimension as the average location of molecules described by that percept. In Figure 3 we show a word cloud for the first dimension of Flavornet and Goodscents in which a word’s size corresponds to this score. For Arctander, which describes molecules with prose, we show the descriptions of the top and bottom five molecules along the first dimension. The first dimension appears to be capturing the same feature in all three datasets – namely pleasantness, which has been utilized in previous work \citep{sobel}\citep{zarz}\citep{kerm}\citep{voss}.

\begin{figure}
  \centering
\includegraphics[scale=.4]{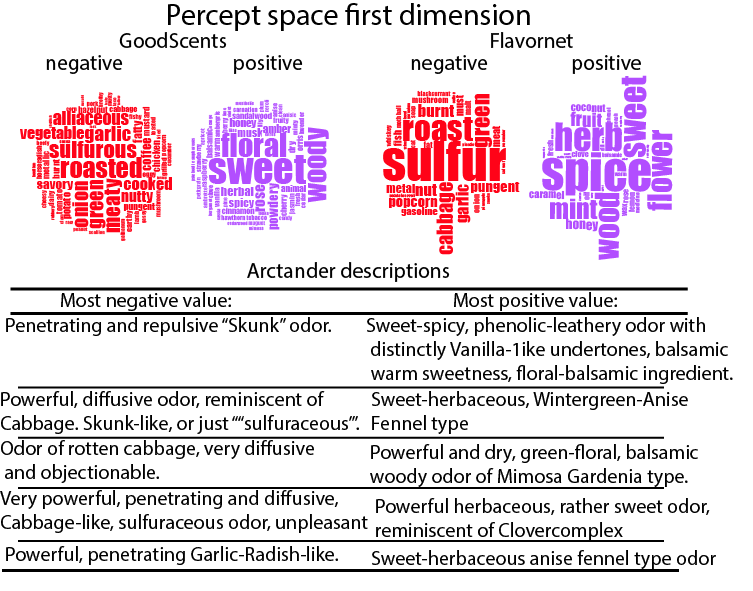}
  \caption{First dimension of perceptual space. Word size in word clouds correspond to the average value of molecules described by that descriptor along the first dimension of perceptual space. Negative values are shown in red, positive in purple. For Arctander, descriptions of the most negative and positive molecules along dimension one are shown.}
\end{figure}

To more objectively relate the 6D spaces to one another, we correlated the percept space distance between molecules which are shared by each pair of datasets. In Figure 4 we show a heatmap of these correlations for all five datasets. Here, we see that local distances between molecules in the three graph spaces are consistent. Additionally, the three graph spaces are consistent with Dravnieks dataset. The Vosshall dataset is distinct, likely for the reasons mentioned previously.  

\begin{figure}
  \centering
\includegraphics[scale=.3]{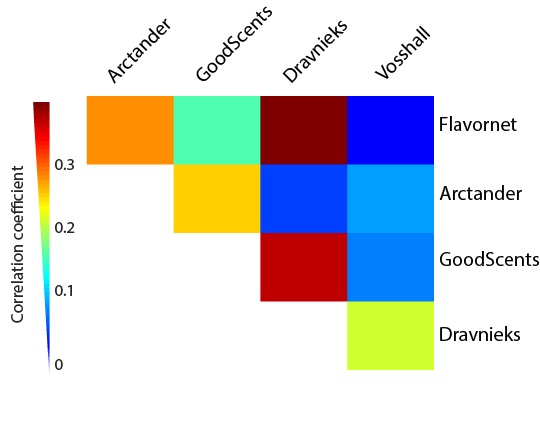}
  \caption{Correlation between datasets. Distances between shared molecules are correlated between each pair of datasets, and the correlation coefficient is represented by color.}
\end{figure}

Using three independent datasets, each with different molecules described by different vocabularies, our method reproduces a 6D space in which local relationships are consistent, and the first dimension corresponds to pleasantness. While this space may be of intrinsic interest to understanding human olfactory perception, we also suggest its utility in predicting olfactory percepts from molecular structure. As we have shown, this 6D space can used to recover individual perceptual features. We therefore propose that predicting a molecule’s olfactory percept from its structure can be rethought of as predicting that molecule’s location in this perceptual space.

\subsection{Predicting percept from molecular structure}

Molecular structures are not directly compatible as input for standard machine learning algorithms. Recently, researchers have been using chemical descriptors, in particular the 4885 Dragon descriptors, to make molecular structures into a machine-learning-compatible input \citep{sobel}\citep{snitz}\citep{voss}. While numerous, these descriptors are highly redundant and are not guaranteed to capture features of structure pertinent for olfactory percept. We devised an alternative, more intuitive method to make predictions from molecular structure.We directly score similarity for each pair of molecules. Using this score to define an inner product space, we embed the molecules. In the resulting space, nearby molecules are molecules that align well. Similar methods have been used in assigning protein function based on amino acid sequence \citep{qiu}. 

\subsubsection{Molecular Alignment Algorithm}

To compute a molecular alignment score between two molecules, we first represented each atom of each molecule as the center of a Gaussian distribution [Figure 5]. The variance of that distribution is the Van der Waals radius of the atom’s element type. The alignment of two molecules is then the overlap of their Gaussian representations. 

We also calculate an alignment score for alignment of each atom type, and like partial charge. For atom scores, an atom was only represented by a Gaussian if it were of the correct element type. Partial charge of each atom was calculated using GAMESS software \citep{gordon}. For a charge alignment score, the amplitude of an atom’s Gaussian representation was the partial charge of that atom. The eight total alignment scores – one for volume, five for atom types (CHONS), and two for partial charge (+/-) – were summed.  

We used simulated annealing to search the space of all shifts and rotations for the optimal alignment between all 3802 molecules - the union of the five datasets. In Figure 5 we show a 3D embedding of the alignment space of the five datasets. By inspection, clusters of molecules in this space indicate that our alignment algorithm is successfully grouping molecules according to their structural similarities. We the suggest the dimensionality of this space is roughly 10 dimensional. With 10 dimensions, the distance between points correlates well with those of the full space.

\begin{figure}
  \centering
\includegraphics[scale=.22]{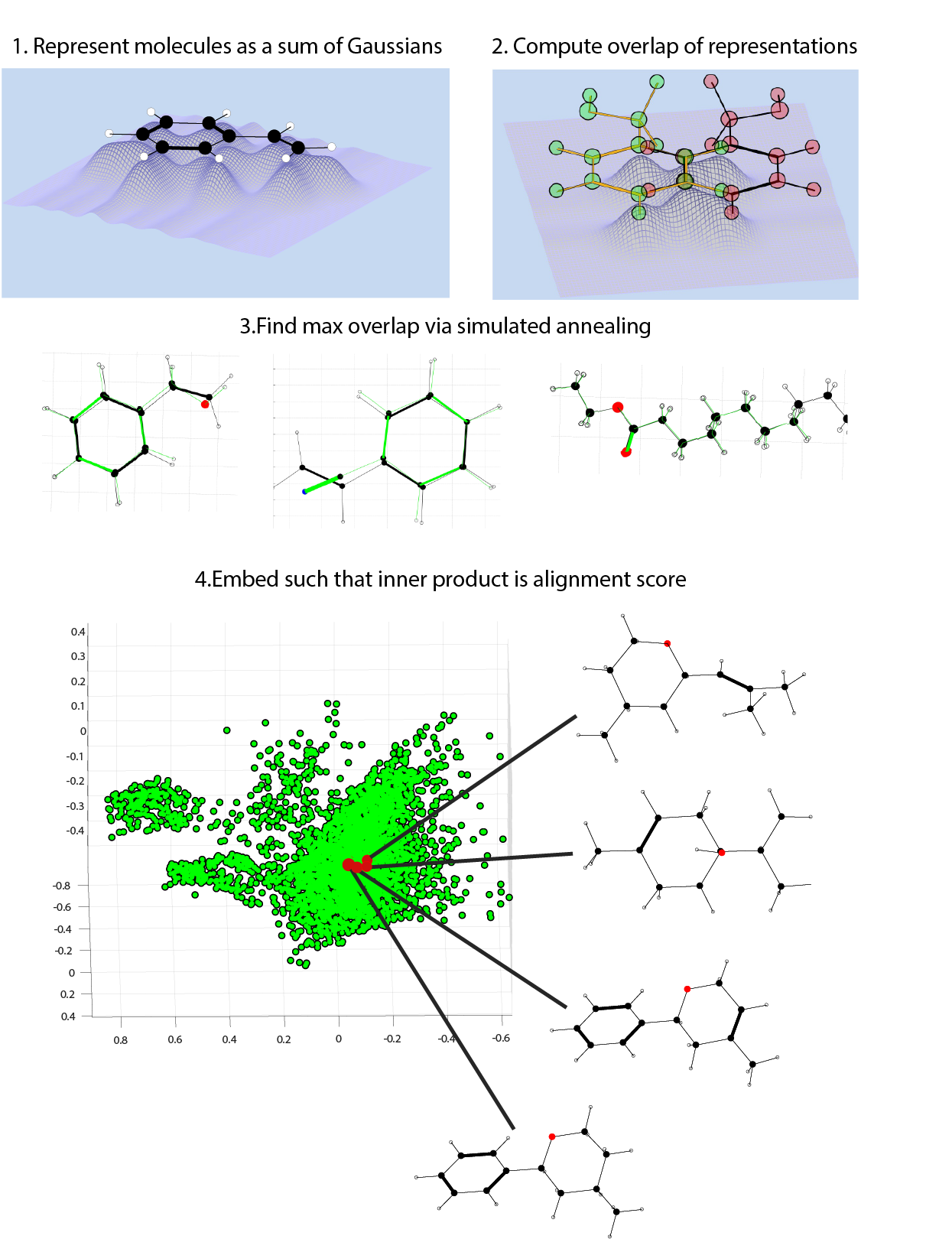}
  \caption{Alignment space algorithm scheme.}
\end{figure}

\subsubsection{Molecular Alignment Kernel Support Vector Regression}

Using support vector regression (SVR), one can make predictions of a continuous variable using only the inner product between data points \citep{cortes}. Here we used the alignment scores as the inner product for support vector regression. We can then avail ourselves of the full structure of the molecules for percept prediction without ever having to explicitly represent the molecular structures with quantitative descriptors.

We compared our alignment space features to the Dragon physicochemical descriptors based on their ability to predict percepts and percept space dimensions ([Figure 6] and [Figure 7]). For both alignment space and Dragon descriptors, we used support vector regression/machines (SVR/SVM) to make predictions. Both used a radial basis function kernel. This allows for SVR and SVM to build nonlinear predictors, with SVR predicting continuous variables and SVM predicting binary classes. Predictions for continuous variables were evaluated by Pearson correlation R, and binary classification were evaluated by their F measure, which is the geometric mean of precision and recall. 

All presented predictions are crossvalidated by a leave-one-out procedure. In this procedure, a classifier or regression is built on the entire dataset spare one molecule. Then, a prediction is made for the spared molecule. This process is iterated, each time sparing a different molecule, until a prediction is made for each molecule with a predictor built without its information. This entire process was also repeated 100 times, and on each iteration the data was resampled with repetition. This gave us a variance estimate for each percept prediction to access significance. In general, the variance of predictions were low, with almost all being on the order of \begin{math} 10^{-4} \end{math}.  

\subsubsection{Alignment algorithm improves individual percept prediction}

First, we made direct predictions of each dataset’s individual percepts. For the Dravnieks and Vosshall dataset, which have continuous ratings, we used support vector regression and Pearson correlation to evaluate performance. For Flavornet and Arctander, we used support vector machines to make binary classifications and used the F measure to evaluate. 

Overall, we found that the 10 dimensional alignment space made significantly better predictions than the 4885 Dragon descriptors on the sparse, binary classifications of Flavornet and Arctander. In particular, in Figure 6, the average F measure of the Dragon descriptors was low, with many individual percepts being predicted zero. For the same percepts, the alignment space produced more generalizable predictions. Even if we compare percepts for which Dragon descriptors produced nonzero F measures, the alignment space still improved binary classification significantly, with a large effect size of 1.4. The same was true for Arctander, in which the effect size for using alignment properties was even larger at 2.3.

For the Dravnieks dataset, the average correlation R from the alignment space predictions are comparable to the predictions made by the Dragon properties [Figure 7]. For many percepts, the alignment properties are again able to find more generalizable relationships, however the Dragon properties perform better on average, with a Cohen’s D effect size of .31. The results from predicting the Vosshall dataset percepts are similar, with a mild Cohen's D effect size of .44 in favor of the Dragon properties. 

\begin{figure}
  \centering
\includegraphics[scale=.20]{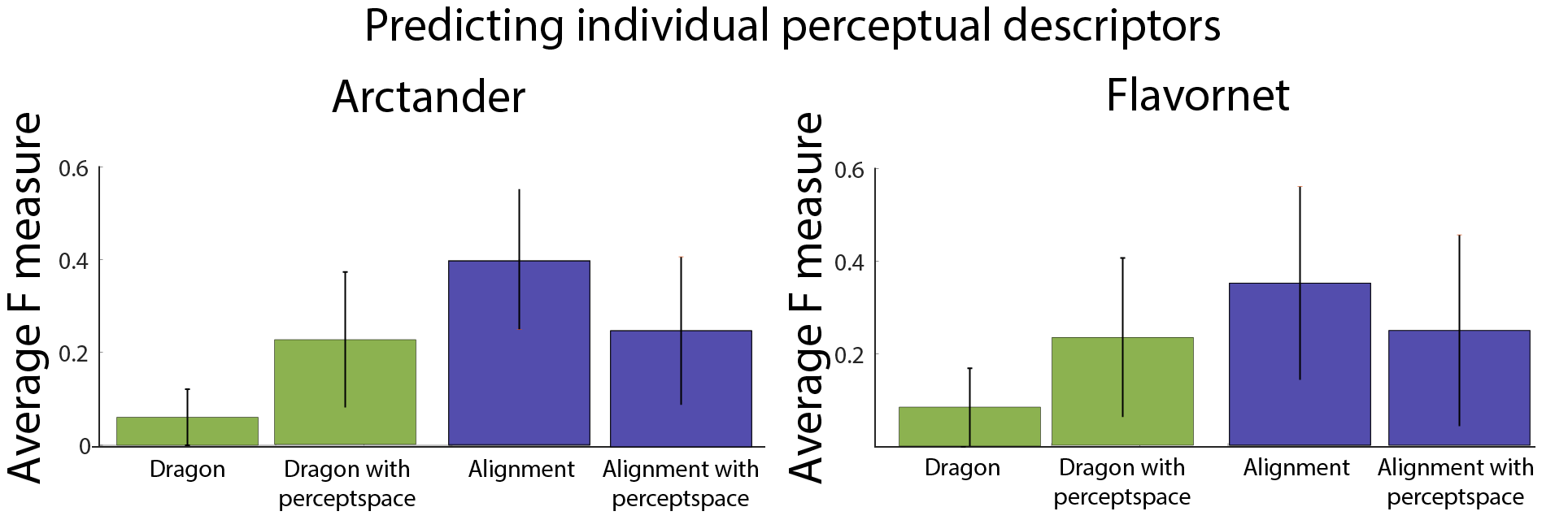}
  \caption{Comparison of Dragon properties with Alignment space on predicting individual perceptual descriptors both directly and using perceptual space.}
\end{figure}

\subsubsection{Percept space dimensions are represented in molecular structure }

We also compared predictions made by alignment space and Dragon properties on predicting the higher order features of the graph embedded percept spaces. 

The transformation from numerous sparse binary predictors to a few continuous dimensions allowed for more robust predictions for both the Dragon and alignment features. Again, we see that the alignment method produces comparable and often improved predictions over the Dragon descriptors. In particular, in Figure 7, one can see that in predicting the first dimension of Flavornet, the putative pleasantness dimension, the alignment space beats out the Dragon descriptors. The bootstrap variance indicates the effect size for this improvement is .63. Overall, however, the effect size for using the alignment space over the Dragon properties for all dimensions is low, .02. For the Arctander dataset, however, predictions are generally better with Dragon properties, with effect size .68 (Figure 7). 

\begin{figure}
  \centering
\includegraphics[scale=.22]{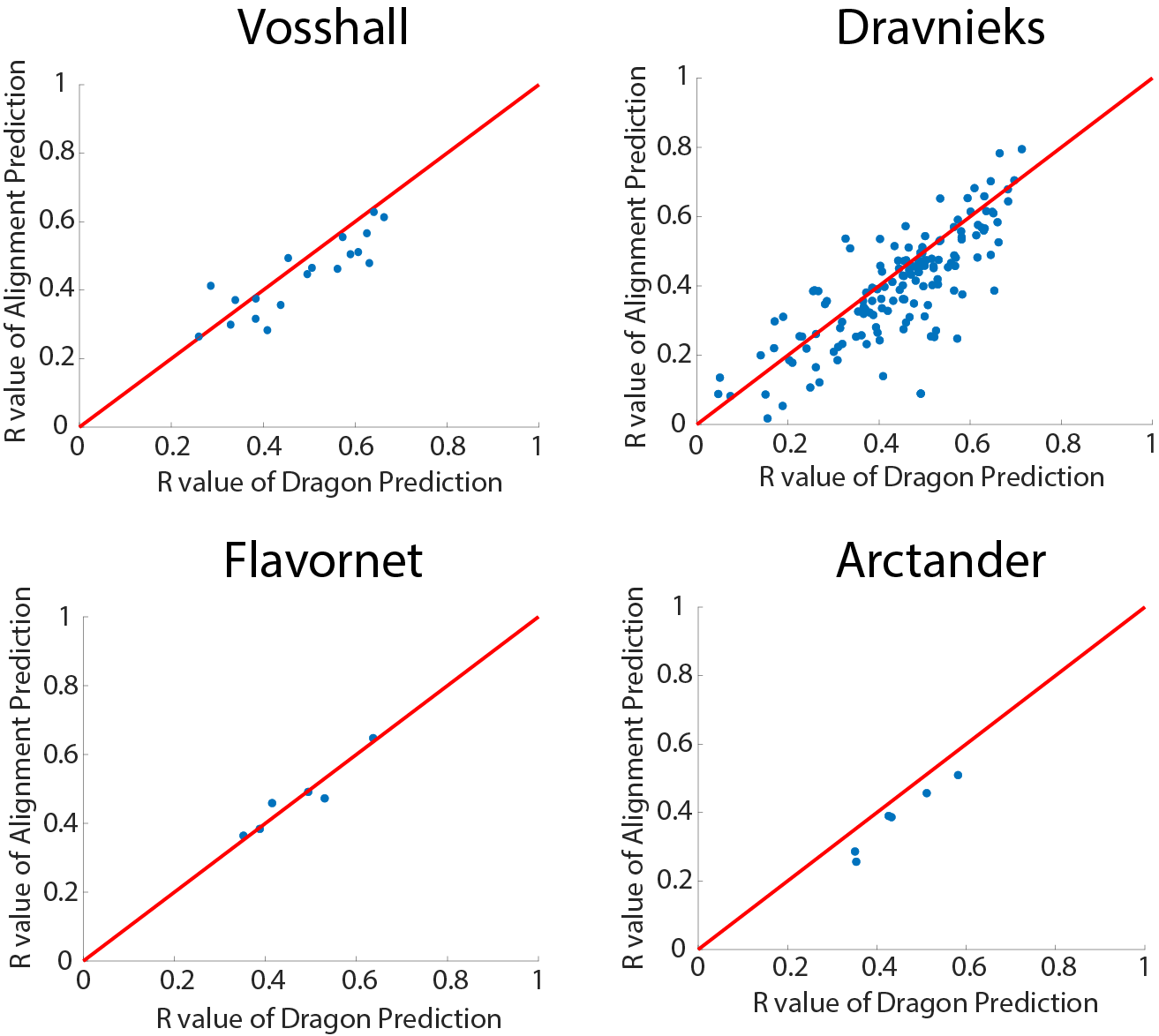}
  \caption{Comparison of Dragon and Alignment features in predicting perceptual space dimensions.}
\end{figure}

\subsubsection{Percept space improves individual percept prediction}

Following our results on the recoverability of individual percepts from the six dimensions of percept space (Figure 2), we used the predicted location of a molecule in 6D percept space to recover individual percepts for each dataset. We show the results of these indirect predictions of individual percepts in Figure 6. Dragon descriptor predictions improved significantly using indirect prediction vs direct, with a large effect size of .88 on Flavornet and 1.13 for Arctander. Interestingly, the alignment space predictions suffered with effect size .49 and .95 in favor of direct predictions for Flavornet and Arctander respectively. This is likely due to the fact that the alignment space’s low dimensional representation already avoids overfitting, and thus the benefit of making indirect predictions from perceptual space is redundant. In comparing the direct alignment predictions to the indirect Dragon predictions, we see the alignment properties still outperform the Dragon properties, with a moderate effect size of .6 for Flavornet and large effect size of 1.15 for Arctander. 

\section{Discussion}

Here we have suggested two independent approaches to improve olfactory percept prediction from molecular structure. The first finding is that sparse qualitative datasets such as Flavornet, GoodScents, and Arctander can be transformed into a 6D continuous manifold in which individual percepts can be readily recovered. These dimensions are intrinsically interesting, appearing to contain the body of human olfactory perception. The first dimension of this space consistently appears as pleasantness, the well-studied dimension used in previous research for olfactory percept prediction. The use of these higher order perceptual features avoids the challenges of sparsely labelled data and enables more robust predictions. Importantly, these predictions can then be used to recover individual percepts with higher fidelity than would be achieved by direct prediction. 

The second finding is that a simple and intuitive approach to capture information from molecular structures is advantageous to the extensive knowledge of the set of Dragon descriptors. Predictions in general are improved for a number of percepts across all five datasets used in this study. In particular, sparsely labelled individual percepts of qualitative datasets are dramatically improved. 

The major appeal of the alignment space is not that it improves predictions, but rather that these predictions come from a parsimonious and intuitive 10D space as opposed to a set of 4885 specialized and often complex physicochemical properties. Furthermore, it is easy to extend the alignment space by including more alignments, or by modifying the weighting of alignment features to emphasize different whole structure features. By contrast, it is difficult to devise a new Dragon property which is not redundant or won’t be lost in the crowd when searching for salient features in a regression. 

 \newpage

\bibliographystyle{unsrt}  
\bibliography{test} 

\begin{thebibliography}{10}

\bibitem{sell}
Sell CS.
\newblock On the unpredictability of odor.
\newblock {\em Angew Chem Int Ed Engl}, 2006.

\bibitem{alex}
Koulakov AA Kolterman BE Enikolopov AG~Rinberg D.
\newblock In search of the structure of human olfactory space.
\newblock {\em Frontiers in systems neuroscience}, 2011.

\bibitem{laska}
Teubner~P Laska~M.
\newblock Olfactory discrimination ability for homologous series of aliphatic
  alcohols and aldehydes.
\newblock {\em Chem Senses}, 1999.

\bibitem{boes}
Lundstrom~JN Boesveldt~S, Olsson~MJ.
\newblock Carbon chain length and the stimulus problem in olfaction.
\newblock {\em Behav Brain Res}, 2010.

\bibitem{joel}
Mainland JD Li YR Zhou T Liu WL~Matsunami H.
\newblock Human olfactory receptor responses to odorants.
\newblock {\em Sci Data}, 2015.

\bibitem{sobel}
Khan RM Luk CH Flinker A Aggarwal A Lapid H Haddad R~Sobel N.
\newblock Predicting odor pleasantness from odorant structure: pleasantness as
  a reflection of the physical world.
\newblock {\em J Neurosci}, 2007.

\bibitem{kerm}
Kermen F Chakirian A Sezille C Joussain P Le Goff G Ziessel A Chastrette M
  Mandairon N Didier A Rouby~C et~al.
\newblock Molecular complexity determines the number of olfactory notes and the
  pleasantness of smells.
\newblock {\em Sci Rep}, 2011.

\bibitem{zarz}
Zarzo M.
\newblock Hedonic judgments of chemical compounds are correlated with molecular
  size.
\newblock {\em Sensors (Basel)}, 2011.

\bibitem{snitz}
"Snitz K Yablonka A Weiss T Frumin I Khan RM~Sobel N".
\newblock " predicting odor perceptual similarity from odor structure.".
\newblock {\em " PLoS Comput Biol"}, "2013".

\bibitem{voss}
Keller A Gerkin RC Guan Y Dhurandhar A Turu G Szalai B Mainland JD Ihara Y Yu
  CW Wolfinger~R et~al.
\newblock Predicting human olfactory perception from chemical features of odor
  molecules.
\newblock {\em Science}, 2017.

\bibitem{dragon}
Mauri A Consonni V Pavan M~Todeschini R.
\newblock Dragon software: An easy approach to molecular descriptor
  calculations.
\newblock {\em Match-Commun Math Co}, 2006.

\bibitem{acree}
Acree TE.
\newblock Gc/olfactometry.
\newblock {\em Anal Chem 69}, 1997.

\bibitem{good}
Luebke W.
\newblock The good scents company.
\newblock {\em Chem Senses}, 1980.

\bibitem{arctander}
Arctander S.
\newblock Perfume and flavor materials of natural origin.
\newblock {\em The Author}, 1960.

\bibitem{dravnieks}
Dravnieks A.
\newblock Atlas of odor character profiles.
\newblock {\em ASTM}, 1985.

\bibitem{iso}
Tenenbaum JB DeSilva V~Langford JC.
\newblock A global geometric framework for nonlinear dimensionality reduction.
\newblock {\em Science}, 2000.

\bibitem{qiu}
Qiu J Hue M Ben-Hur A Vert JP~Noble WS.
\newblock A structural alignment kernel for protein structures.
\newblock {\em Bioinformatics}, 2007.

\bibitem{gordon}
Gordon MS Alexeev Y Gan ZT Schmidt MW Kendall RA Ivanic J~Ruedenberg K.
\newblock New advances in scalable electronic structure methods.
\newblock {\em Abstr Pap Am Chem S}, 2002.

\bibitem{cortes}
V~Cortes~C, Vapnik.
\newblock Support vector networks.
\newblock {\em Machine Learnings}, 1995.

\end{thebibliography}

\end{document}